\documentclass[11pt,letter,superscriptaddress,prb,aps]{revtex4-1}
\usepackage[utf8x]{inputenc}
\usepackage{amsmath,amssymb}
\usepackage{graphicx}% Include figure files
\usepackage{dcolumn}% Align table columns on decimal point
\usepackage{bm}% bold math
\usepackage{subfigure}
\usepackage{enumerate}
\usepackage{multirow}
\usepackage{tabularx}
\usepackage{threeparttable}
\usepackage{epstopdf}
\usepackage{paralist}
\usepackage{lineno}
\usepackage{color}
\usepackage{soul}
\sethlcolor{yellow}

\newcommand{\ket}[1]{\left| #1 \right\rangle}
\newcommand{\bra}[1]{\left\langle #1 \right|}

\newcommand{\expn}[1]{{\rm e}^{#1}}
\newcommand{\bv}[1]{\mathbf{#1}}

\newcommand{\COMMENT}[1]{}
\newcommand{\textapprox}{{\raise.17ex\hbox{$\scriptstyle\mathtt{\sim}$}}}
\begin{document}

 %\linenumbers
 %\begin{linenumbers}

 \title{Substantial optical dielectric enhancement by volume compression in LiAsSe$_2$}
 
 \author{Fan Zheng}
  \affiliation{The Makineni Theoretical Laboratories, Department of Chemistry, University of Pennsylvania, Philadelphia, Pennsylvania  19104-6323, USA}
 \author{John A. Brehm}
  \affiliation{The Makineni Theoretical Laboratories, Department of Chemistry, University of Pennsylvania, Philadelphia, Pennsylvania  19104-6323, USA}
 \author{Steve M. Young}
   \affiliation{Center for Computational Materials Science, Naval Research Laboratory, Washington, D.C. 20375, USA}
 \author{Youngkuk Kim}
  \affiliation{The Makineni Theoretical Laboratories, Department of Chemistry, University of Pennsylvania, Philadelphia, Pennsylvania  19104-6323, USA}

 \author{Andrew M. Rappe}
 \affiliation{The Makineni Theoretical Laboratories, Department of Chemistry, University of Pennsylvania, Philadelphia, Pennsylvania  19104-6323, USA}

 \begin{abstract}
  Based on first-principles calculations, we predict a substantial increase in the optical dielectric function of LiAsSe$_2$ under pressure. We find that the optical dielectric constant is enhanced threefold under volume compression. This enhancement is mainly due to the dimerization strength reduction of the one-dimensional (1D) As--Se chains in LiAsSe$_2$, which significantly alters the wavefunction phase mismatch between two neighboring chains and changes the transition intensity. By developing a tight-binding model of the interacting 1D chains, the essential features of the low-energy electronic structure of LiAsSe$_2$ are captured. Our findings are important for understanding the fundamental physics of LiAsSe$_2$ and provide a feasible way to enhance the material optical response that can be applied to light harvesting for energy applications.
 \end{abstract}

 \maketitle 
 \newpage

   \section{Introduction}

   The dielectric response, as a fundamental physical property of materials, describes how materials respond to an external electric field. In semiconductors, when the applied electric field frequency is in the range of visible light, the photon excitation of electronic inter-band transitions dominates the total dielectric response, which is described by the optical dielectric function. The optical dielectric function is strongly related to other optical properties of the material, including light absorption, refraction and non-linear optical responses. Therefore, the enhancement and tunability of the optical dielectric function of a material are significantly important in various areas, such as solar cell, optical devices and sensors. A great deal of research has been done to increase the material optical dielectric response. In particular, defects, material doping and surface plasmon induced by metallic nanoparticles have been widely used to increase the optical absorption in semiconductors~\cite{Chikamatsu04p127,Schaadt05p063106,Beck09p114310,Santra12p2508,Li09p8460}. Whereas most of the previous methods rely on the assistance of another material, the intrinsic bulk dielectric response enhancement of the light absorber is less studied. 
   
   Alkali-metal chalcogenides such as KPSe$_6$, K$_2$P$_2$Se$_6$, LiAsSe$_2$, LiAsS$_2$ and NaAsSe$_2$ have been synthesized, and their band gaps lie in the visible light region~\cite{Bera10p3484}. Since they have spontaneous polarization, these materials are potential candidates to show the bulk photovoltaic effect~\cite{Brehm14p204704}. Moreover, strong optical second-harmonic generation susceptibility has been observed experimentally and theoretically~\cite{Bera10p3484,Song09p245203}. However, the effect of structural distortion on their linear optical responses has not been studied, and the structure-property-optical performance relationship is still unclear. In this paper, by using a first-principles method, we show that the optical dielectric constant of LiAsSe$_2$ increases threefold by volume compression. More interestingly, As and Se atoms in LiAsSe$_2$ form weakly interacting quasi-one-dimensional atomic chains, of which the dimerization strength can be tuned by volume compression. Atomic chains have attracted a great deal of interest, due to their one-dimensional nature giving rise to exotic phenomena such as conductivity~\cite{Cretu13p3487,Wang15p020508}, metal-insulator transition~\cite{LaTorre15p6636}, and topological phases~\cite{Farhan14p042201,Steinberg14p036403}. Herein, their important roles in light absorption are emphasized. As illustrated by a tight-binding model, the dimerization strength is strongly coupled to the relative phases of the gap state wavefunctions between the two neighboring chains. By reducing the wavefunction phase mismatch between the chains, the magnitude of transition intensity for the transitions near the band edges increase significantly, giving rise to substantial optical dielectric function enhancement.

   \section{Computational Details}
\begin{figure}[h]
    \includegraphics[width=5in]{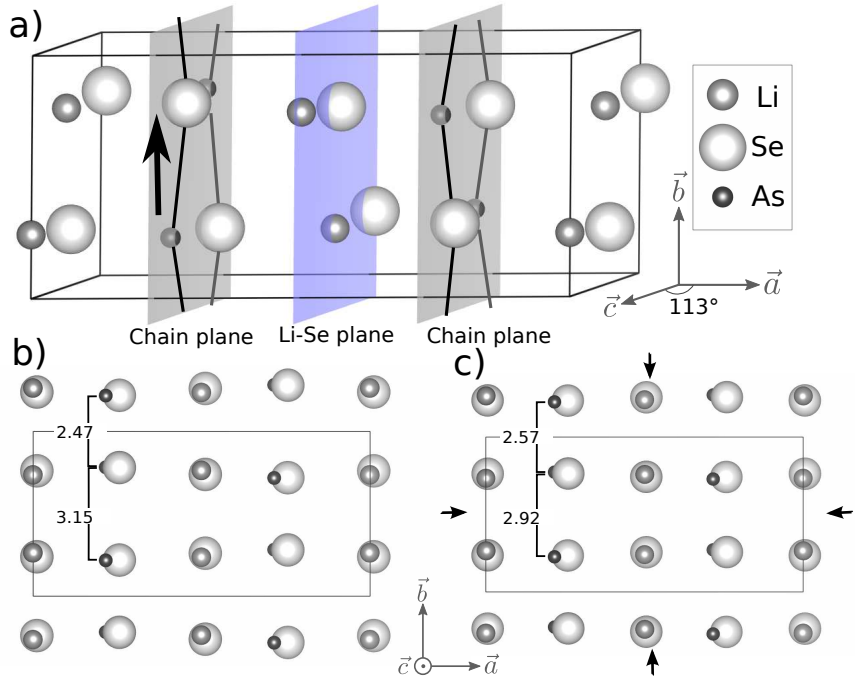}
    \caption{a) The unit cell of LiAsSe$_2$. The lines between As and Se atoms indicate the quasi-one-dimensional chains. The chain with its neighbor chains form a chain plane (grey color plane). These parallel chain planes are separated by the Li-Se plane (light purple plane) in the middle. b) Side view of the experimental structure (ES). c) Side view of the compressed structure (CS). The differences between the ES and the CS are mainly ion motions in the $\vec{b}$ direction. As illustrated by the bond lengths between two neighboring As--Se bonds, ES shows stronger dimerization strength along the chain than CS.}\label{structure}
   \end{figure}

   Figures~\ref{structure}a and b show the experimental structure (ES) of LiAsSe$_2$~\cite{Bera10p3484}. The polar phase of LiAsSe$_2$ has the $Cc$ space group with the glide plane perpendicular to the lattice vector $\vec{b}$. The polarization induced by ionic displacement lies in the $\vec{a}$-$\vec{c}$ plane~\cite{Brehm14p204704}. As shown, the As and Se atoms form distorted quasi-one-dimensional atomic chains along the $\vec{b}$ direction. This chain and its neighboring chains form a two-dimensional chain plane (illustrated as the grey plane), and these chain planes are separated by Li-Se planes (light purple plane). In the ES, this As--Se chain dimerizes, creating alternating As--Se bonds with two different bond lengths.

   The plane-wave density functional theory (DFT) package QUANTUM-ESPRESSO was used to perform structural relaxations and electronic structure calculations, with the Perdew-Burke-Ernzerhof (PBE) generalized gradient approximation exchange-correlation functional~\cite{Giannozzi09p395502}. Norm-conserving, designed non-local pseudopotentials were generated with the OPIUM package~\cite{Rappe90p1227,Ramer99p12471}. A plane-wave cutoff energy of 50 Ry was sufficient to converge the total energy with the $k$-point sampling on a 4$\times$8$\times$8 grid. The structure relaxed with the PBE functional underestimates the dimerization along the chain, and it does not match with the ES. By using the GGA + $U$ method with effective Hubbard $U_{\rm{eff}} = 7.5$ eV on the As 4 $p$ orbitals, the relaxed structure matches the ES very well. Adding $U$ on $p$ orbitals to get the correct structure is not rare, as the large self-interaction error originating from $s$ or $p$ orbitals may partially be corrected by the DFT+$U$ method~\cite{Erhart14p035204,Slipukhina11p137203}. The DFT calculated band gap is 0.8 eV, which underestimates the experimentally measured 1.1 eV~\cite{Bera10p3484}. With the converged charge density, the wavefunctions used for the dielectric function calculations are obtained from non-self-consistent calculations performed on a denser $k$-point grid with a sufficient number of empty bands. By using the long wavelength approximation and the single particle approximation, the imaginary part of the optical dielectric function is calculated as Eq.\eqref{eq:dielectric},
   
   \begin{flalign}\label{eq:dielectric}
      \epsilon_{2,ii}(\omega) &= \frac{\pi}{2\epsilon_0}\frac{e^2}{m^2\left(2\pi\right)^4\hbar\omega^2}\sum_{c,v}\int_{BZ} d\bv{k} \left| \bra{c,\bv{k}}p_i\ket{v,\bv{k}}\right|^2 \delta(\omega_{c,\bv{k}}-\omega_{v,\bv{k}}-\omega) 
   \end{flalign}
   
   \noindent where $\omega$ is the light frequency; $i$ is the Cartesian coordinate; $\bv{k}$ is the Bloch wave vector; $c$, $v$ denote the conduction and valence band with energy $\hbar\omega_{c/v}$. The real part of the dielectric function, $\epsilon_1$, can be calculated from the Kramers-Kronig relation.

   \section{Results and Discussion}
\begin{figure}
    \includegraphics[width=4.5in]{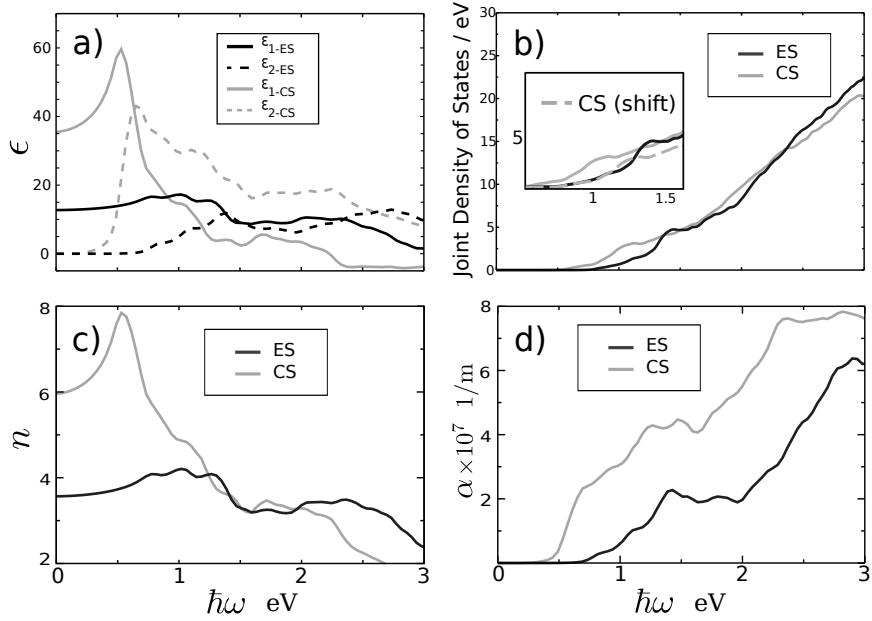}
    \caption{a) LiAsSe$_2$ optical dielectric ($\epsilon$) function spectrum of the ES and CS as a function of photon energy along $\vec{b}$. $\epsilon_1$ is the real part of the dielectric response spectrum, and $\epsilon_2$ is the imaginary part. b) Joint density of states for the two structures. Owing to the different band gaps of the ES and CS (0.2 eV difference), the inset graph shows the shifted-CS (shifting the spectrum by 0.2 eV) and ES JDOS spectrums in order to compare with the same band gaps. c) Refractive index ($n$) spectrum along $\vec{b}$. d) Absorption coefficient ($\alpha$) spectrum $\vec{b}$. }\label{dielectric_spec}
   \end{figure}

    The compressed structure (CS), with much weaker dimerization strength of the atomic chains (Fig.~\ref{structure}c), is obtained by compressing all the lattice vectors by 3\%, followed by the relaxation of the internal atomic positions. This volume compression of LiAsSe$_2$ strongly enhances its optical dielectric response as shown in the calculated optical dielectric functions of the ES and CS (Fig.~\ref{dielectric_spec}a). Figures~\ref{dielectric_spec}b, c and d illustrate the calculated joint density of states (JDOS), refractive index and absorption spectrum along the $\vec{b}$ direction as a function of the photon energy for the ES and CS, respectively. As shown from the spectrum, the CS shows much higher linear optical responses near the band gap than the ES. In particular, the optical dielectric constant of the CS increases to more than three times its original value (Fig.~\ref{dielectric_spec}a). The imaginary part of the optical dielectric function, describing the real electronic inter-band transitions, also shows great enhancement under compression. As expressed in Equation~\eqref{eq:dielectric}, the imaginary part of the dielectric function is the product of JDOS $\sum_{c,v,\bv{k}}\delta (\omega_{c,\bv{k}}-\omega_{v,\bv{k}}-\omega)$ and the transition intensity $\left| \bra{c,\bv{k}}p_i\ket{v,\bv{k}}\right|^2$. However, we find that the JDOS contribution to the enhancement is negligible. As shown in Fig.~\ref{dielectric_spec}b, in the energy range $0<\hbar\omega<2$ eV where the imaginary part $\epsilon_2$ shows substantial enhancement, the calculated JDOS for the ES and CS have very similar magnitude. Therefore, this dielectric function enhancement mainly comes from the increase of transition intensity by compression.

\begin{figure}[h]
    \includegraphics[width=6in]{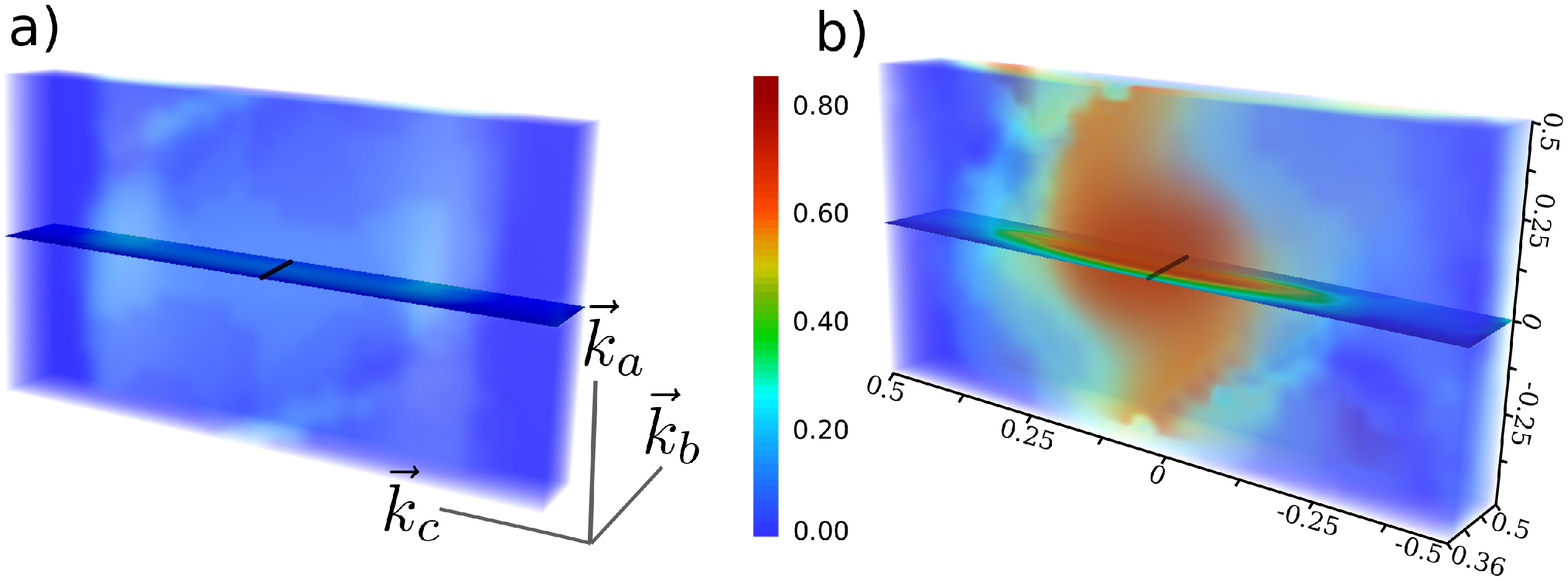}
    \caption{Distribution of $\left|\bra{\psi_{v,\bv{k}}} \bv{p} \ket{\psi_{c,\bv{k}}}\right|^2/V$ (eV/\AA$^3$) in the Brillouin zone (BZ) extracted from DFT calculation of LiAsSe$_2$ for a) the ES and b) CS. $V$ is the volume of the unit cell. For simplicity, the primitive BZ is illustrated as an orthogonal box with reciprocal lattice vectors $\vec{k}_a$, $\vec{k}_b$, and $\vec{k}_c$ along the three edges of the box. The transitions with transition energy less than 2 eV are plotted, as this energy region shows the greatest dielectric function enhancement. The detailed transition intensity profiles along the black lines in the figures for the ES and CS are shown in Fig.~\ref{tb_dipole}b.}\label{3d_distr}
   \end{figure}

\begin{figure}[t]
    \includegraphics[width=6.22in]{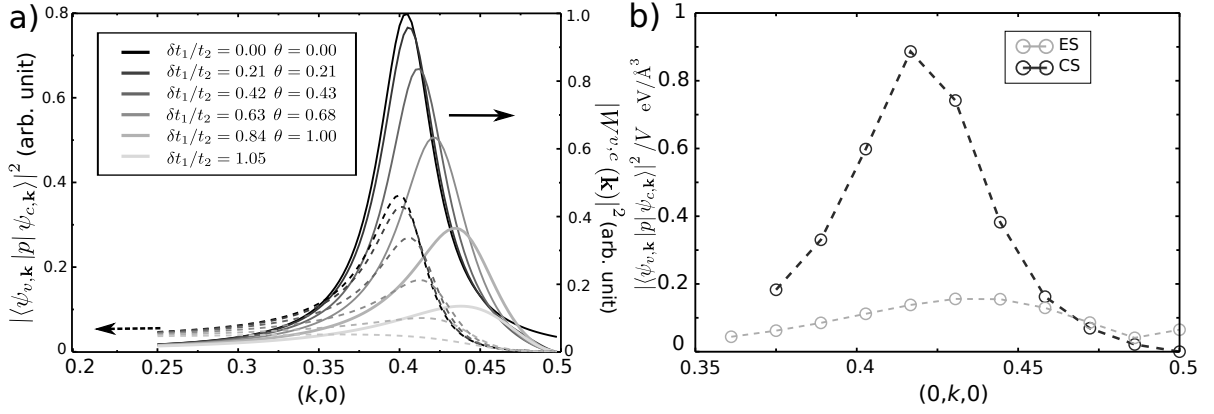}
    \caption{a) Calculated transition intensities $\mathcal{I}(\bv{k})$ and $\left|W^{v,c} \left( \bv{k} \right)\right|^2$ ($W^{v,c}\left(\bv{k}\right) = \sum_{j,j'} C_{j',\bv{k}}^{v \ast} C_{j,\bv{k}}^{c}$) from 2D TB model. The $x$ axis is the wavevector along the chain direction. b)  Transition intensities (eV/\AA$^3$) extracted from DFT calculation of transition intensity for the ES and CS. They are plotted along the chain direction as indicated by the short black lines in Fig.~\ref{3d_distr}.}\label{tb_dipole}
   \end{figure}

    To further resolve the origin of the dielectric enhancement by pressure, we present the distribution of the transition intensity as a function of $\bv{k}$ in momentum space. Figure~\ref{3d_distr} shows the transition intensity distributions in the Brillouin zone (BZ), with the transitions between the valence band maximum (VBM) to the conduction band mimimum (CBM) within the the energy range of 0--2 eV, since these transitions are dominant in the dielectric function enhancement. As displayed in Figure~\ref{3d_distr}, the $k$-resolved distributions show distinct patterns in addition to their overall differences in the corresponding dielectric constants. For the ES, most of the $\bv{k}$ points have similar yet low magnitude of transition intensities. However, for the CS, the $k$-resolved transition intensity shows significant changes, with the high magnitude $\bv{k}$ points mostly distributed on a thin plane perpendicular to the reciprocal lattice vector $\vec{k}_b$. The $\bv{k}$ points contributing the highest transition intensities are broadly located in the middle region in this plane. Along the $\vec{k}_a$ and $\vec{k}_c$ directions of this plane, the transition intensity changes slowly with respect to wavevectors, indicating the weak bonding character. This can be attributed to the weak As/Se--Se/Li inter-planar and As--Se inter-chain interactions. However, the magnitude of transition intensity shows rapid change along the $\vec{k}_b$ direction, as illustrated by the transition intensity profile along this direction (Fig.~\ref{tb_dipole}b). This strong $k$-dependent transition intensity distribution reveals the strong covalent bonds character along the chain direction. The highly inhomogeneous distribution of the transition intensity can be considered as an indication of the quasi one-dimensional nature of the system near the low-energy spectrum, steming from the dimerization changes of the As and Se atoms. Therefore, investigating the electronic structure of the chains is essential to further understand the origin of the dielectric response enhancement. 
    
\begin{figure}[h]
    \includegraphics[width=6in]{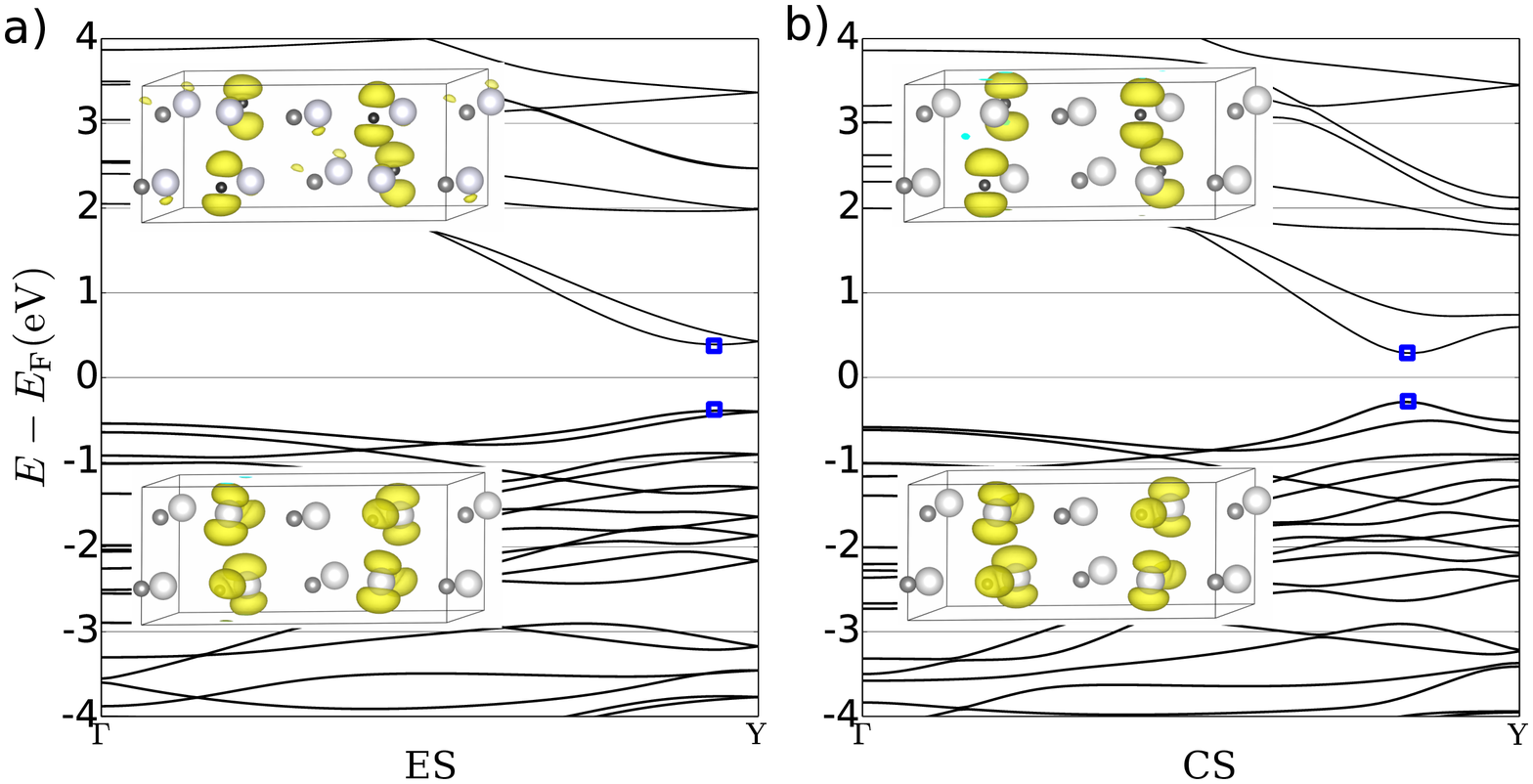}
    \caption{The band structure of LiAsSe$_2$ from $\Gamma$ to Y $\left(0,\pi/b,0\right)$ along $\vec{k}_b$, and the charge density isosurfaces of the conduction band minimum (CBM) and valence band maximum (VBM) states indicated by the blue squares in the band structures for a) ES and b) CS.}\label{bands}
   \end{figure}

   Figure~\ref{bands} shows the DFT band structures plotted along $\Gamma$--Y. Under compression, most bands along the $\vec{k}_b$ direction show relatively small changes, except the bands near the band edges. The CS shows strong dispersion near its optical gap at $\left(0.0, 0.42,0.0\right)$ (fractional coordinate), as its CBM shows a ``dip" while the VBM shows a ``bump". The ES has its band gap shifted towards the BZ boundary at $\left(0.0, 0.46,0.0\right)$. Comparing to those of the CS, the bands of the ES near the band gap shows much less dispersion, and the dip and bump features become less obvious. Besides the band dispersion change, the band gap shows noticeable change from 0.80 eV (ES) to 0.62 eV (CS). More importantly, we find that the inter-band transition between the band edges in the CS provides the highest transition intensity magnitude, but this corresponding value in the ES is very low. In order to understand the bonding characters of these states which give the highest transition intensity, the charge density isosurfaces of the VBM and CBM are plotted in Figure~\ref{bands}. Unexpectedly, both the ES and CS show quite similar charge density distributions, with non-bonding Se $p$ orbital character as VBMs and non-bonding As $p$ as CBMs, suggesting that the atomic orbital overlaps cannot explain such large dielectric enhancement by compression due to their similar charge densities. Rather, we find that the dimerization change induced by the compression can strongly alter the phase of wavefunctions so as to vary transition intensity magnitude significantly, as we will discuss below.
   
\begin{figure}
    \includegraphics[width=5.8in]{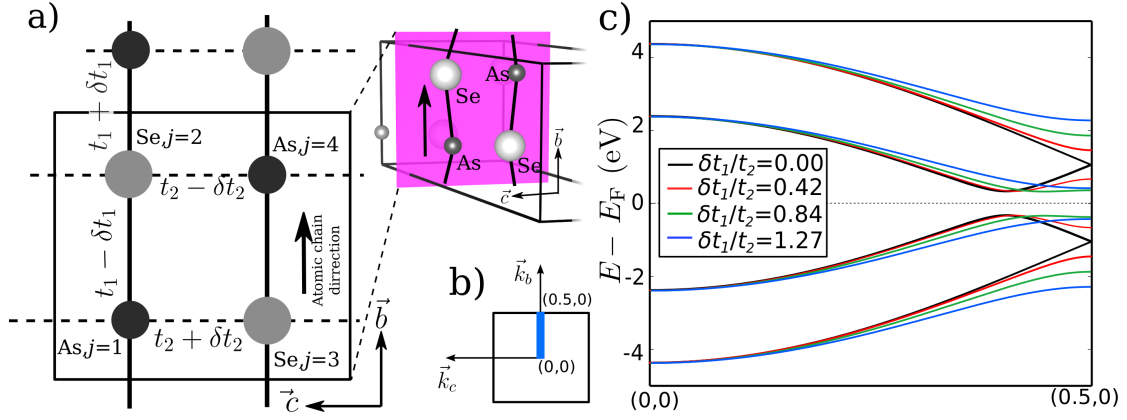}
    \caption{a) 2D TB model for weakly interacting As--Se chains (the inset graph shows the chains in LiAsSe$_2$). The dashed lines indicate the chain-chain interaction connecting the As--Se chains (solid lines). $t\pm\delta t$ denotes the hopping strength. b) The Brillouin zone of the 2D model. The band structure (graph c) is plotted along the thick blue line. c) The band structure calculated from the 2D TB model along the chain propagation direction under different dimerization strengths ($\delta t_1/t_2$ with $t_2$ fixed).}\label{tb_bands}
   \end{figure}

   To demonstrate the significant influence of wavefunction phase change on the optical response enhancement, we construct a two-dimensional (2D) tight-binding (TB) model with interacting atomic chains illustrated in Figure~\ref{tb_bands}a. The TB model comprises four orbitals ($j=$1, 2, 3, 4) in a square lattice with lattice constant $a$ and periodic boundary conditions along $\vec{b}$ and $\vec{c}$ to model a chain plane in LiAsSe$_2$. Owing to the weak interaction between the chain plane and the Se--Li plane, the inter-planar interaction along the $\vec{a}$ direction is not considered. As shown in the charge density distributions (Fig.~\ref{3d_distr}), the $p$ orbitals from the As and Se atoms form $\sigma$-type covalent bonds along the chain (along the $\vec{b}$ direction). The dimerized hopping strength is denoted as $t_1 \pm \delta t_1$ to describe the alternating As--Se bond lengths. Across the chains ($\vec{c}$-direction), $\pi$-bonding between the $p$ orbitals forms, where the corresponding hopping interaction is denoted as $t_2 \pm \delta t_2$. We find that this inter-chain interaction is of crucial importance in reproducing the correct DFT band structure, although these interactions are weak relative to the intra-chain interaction, thus assuming $\left|t_2\right| < \left|t_1\right|$. The onsite energies of As and Se orbitals are set to $E_0 + \delta E$ and $E_0 - \delta E$, respectively. The onsite energies and hopping strengths of the TB Hamiltonian are tuned to reproduce the DFT band structure.
   
   By solving the TB model numerically, we obtain the band structures in Fig.~\ref{tb_bands}c plotted along the chain propagation direction indicated by the blue line in Fig.~\ref{tb_bands}b. We also calculate the band structures by gradually reducing the dimerization strength (decreasing $\delta t_1$) with $t_2$ fixed, and find that the band gap position shifts away from the BZ boundary. More interestingly, the dispersion of the band edges are significantly enhanced when decreasing $\delta t_1$. This feature becomes clearer by calculating the $k$-resolved transition intensity using the TB wavefunctions.
   
   As derived in the Appendix, the transition intensity ($\mathcal{I}$) is expressed as $\mathcal{I}\left(\bv{k}\right)=\left|W^{v,c} \left( \bv{k} \right) \Pi\left(\bv{k}\right)\right|^2$, with $W^{v,c}=\sum_{j,j'}C_{j',\bv{k}}^{v \ast} C_{j,\bv{k}}^{c}$, summing over the contributions of the wavefunction coefficients to the transition intensity. In this case, $\Pi$ accounts for the contribution generated when constructing the TB basis set from the localized atomic orbitals, with its relative value only determined by the wavevector without solving the Hamiltonian. Shown in Fig.~\ref{tb_dipole} is the calculated $\mathcal{I}$ along the same $k$-path as used in the TB band structure (Fig.~\ref{tb_bands}b). By reducing the dimerization strength of the atomic chains (reducing $\delta t_1/t_2$), the transition intensities for the band edge states and nearby increase significantly, which agrees well with the DFT transition intensity trend under compression shown in Fig.~\ref{tb_dipole}b. Additionally, by plotting $W$ which is contributed only from the wavefunction as we are interested, it is clear that it shows exactly the same trend as $\mathcal{I}$, demonstrating the significant role of wavefunctions in the enhancement of the transition intensity under pressure. 
   
   The low-energy $k\cdot p$ effective theory provides simpler and more explicit band structure and wavefunction expression. The Hamiltonian $H\left(\bv{k}\right)$ is further expanded in the vicinity  of the BZ boundary as $H(\bv{k})=H(\bv{K})+(\bv{k}-\bv{K})H'(\bv{K})$ with ${\bv{k}} = {\bv{K}} + (q, 0)$, $\bv{K} = \left(\pi/b,0\right)$. From the $k\cdot p$ Hamiltonian, the energies for the valence band ($E_{-}$) and the conduction band ($E_{+}$) near the BZ boundary are obtained as:
   
   \begin{flalign}
    E_{\pm}\left(q\right) &= \pm \sqrt{\delta E^2 - 2 \sqrt{4t_2^2\Omega\left(q\right)} + 4\delta t_2^2 + \Omega\left(q\right)} \\
    \Omega\left(q\right)  &= 4\delta t_1^2 + \left(q a t_1\right)^2
   \end{flalign}
   
   \noindent When $\left|\delta t_1\right| > \left|t_2\right|$, the band gap is at the BZ boundary ($q = 0$). By decreasing the dimerization strength such as $\left|\delta t_1\right| < \left|t_2\right|$, the band gap wavevector ($q(E_{\rm g})$) is $2\sqrt{t_2^2 - \delta t_1^2}/\left(a t_1\right)$. This change of band gap position as a function of the dimerization strength ($\delta t_1/t_2$) agrees with our DFT band structures of LiAsSe$_2$. In the ES, the strong dimerization between the As and Se atoms moves the band gap close to the BZ boundary. In the CS, the reduced dimerization due to the compression shifts the band gap away from the boundary, giving rise to the strong dispersion for the states near the gap. 
   
   Within this low energy theory, the phase relationships of the wavefunctions are further explored by evaluating the analytical expression of wavefunctions for the band edge states. When the band gap is not at the BZ boundary ($\left|\delta t_1\right| < \left|t_2\right|$) , the wavefunctions of the gap states have simple forms:
   
   \begin{flalign}
    \psi_{\rm VBM} &= 1/\sqrt{2}\left(0,\expn{i\theta},1,0\right),  \\
    \psi_{\rm CBM} &= 1/\sqrt{2}\left(1,0,0,\expn{i\theta}\right)
   \end{flalign}\label{edge}
   
   \noindent where the wavefunctions are written with the TB basis of the four orbitals: $\chi_{{\rm As},j=1}$, $\chi_{{\rm Se},j=2}$, $\chi_{{\rm Se},j=3}$, and $\chi_{{\rm As},j=4}$ (Fig.~\ref{tb_bands}a). Due to the simple form of the wavefunctions, we use these two states to show the effect of phases of the chains. Here, $\theta = \arcsin\left(\left|\delta t_1 /t_2\right|\right)$ indicates the dimerization strength of the atomic chains. From the wavefunction expression, it is clear that the VBM and CBM are always non-bonding states without mixing of the As and Se orbitals, which is also observed in the DFT calculation. More interestingly, $\theta$ controls the phase mismatch between the wavefunctions of the two chains in the chain plane. For example, for the CBM wavefunction, when $\theta=0$, the orbitals on $\chi_{{\rm As},j=1}$ and $\chi_{{\rm As},j=4}$ are populated in the same phase, while, with nonzero $\theta \ne 0$, $\chi_{{\rm As},j=1}$ on one chain and $\chi_{{\rm As},j=4}$ on the neighboring chain have the phase difference of $\expn{i\theta}$ between the corresponding wavefunction coefficients. Hence, the application of the hydrostatic stress to LiAsSe$_2$ reduces $\theta$, enabling wavefunction phase matching between the two neighboring atomic chains, which is essential to the enhancement of dielectric responses.
   
   Using this simple form of wavefunctions, the band edge transition intensity is evaluated as $\mathcal{I}\propto\left|\expn{i\theta}+\expn{-i\theta}\right|^2=\cos^2\left(\theta\right)=1-\left(\delta t_1/t_2\right)^2$. The first exponential term $\expn{i\theta}$ originates from one chain and the other term from the neighboring chain. In this material, the finite dimerization strength of the two neighboring chains have opposite effects to their contributions to the transition intensity. Therefore, without varying the overlaps between the atomic orbitals, the phase change of the wavefunction induced by structural change alters the overall dielectric function significantly.

  \section{Conclusion}

   In summary, by using a first-principles method, we have shown that volume compression can significantly enhance the optical dielectric function and the dielectric constant by factor of three in LiAsSe$_2$. This material is essentially a network of As--Se 1D atomic chains with the dimerization strength tunable by compression. The enhancement of the transition intensity near band edges is the main reason of the overall dielectric function improvement. A 2D tight-binding model with weakly interacting atomic chains is developed to explore the relation of dimerization strength and transition intensity. When the dimerization is strong, the wavefunctions of the two neighboring chains have significant phase mismatch, providing destructive interference that reduces to the dielectric function. By reducing this wavefunction phase mismatch via compression, the collective contributions from the chains dramatically enhance the overall dielectric response and light absorption. Our results indicate that this material is suitable as the light absorber in the solar cell application. Furthermore, since the transition intensity is related to other optical processes such as second-harmonic generation and the non-linear optical effects, we expect that the volume compression can enhance their responses.

  \section{Acknowledgment}
   F.Z. was supported by the Department of Energy under grant DE-FG02-07ER46431. J.A.B. was supported by the National Science Foundation under grant DMR-1124696. S.M.Y. was supported by the Office of Naval Research under grant N00014-11-1-0664. Y.K. was supported by the National Science Foundation under grant DMR-1120901. A.M.R. was supported by the Office of Naval Research under grant N00014-12-1-1033. The authors acknowledge computational support from the HPCMO of the DOD and the NERSC center of the DOE.
   \\
  
\appendix*

{\bf APPENDIX}

The Bloch wavefunction based on the TB orbitals is:

\begin{flalign}
 \psi_{n,\bv{k}} = \sum_{j} C_j^{n,\bv{k}} \chi_j^{\bv{k}}
\end{flalign}

\noindent $\chi_{j}^{\bv{k}}$ ($j$=1, 2, 3, and 4) is expanded as $\sum_{\bv{R}}\expn{i\bv{k}\cdot \left( \bv{R}+\bv{s}_j \right)} \phi_{\bv{R},j}$, and $\phi_{\bv{R},j}$ is the localized atomic orbital centering at the position of $\bv{R}+\bv{s}_j$.

With the Bloch wavefunctions, the transition intensity is expressed as:

\begin{flalign}
 \mathcal{I}(\bv{k}) &= \left|\bra{\psi_{v,\bv{k}}} \bv{p} \ket{\psi_{c,\bv{k}}}\right|^2 \nonumber\\
                     &= \left|\sum_{j,j'} C_{j',\bv{k}}^{v \ast} C_{j,\bv{k}}^{c} \Pi_{jj',x}\left(\bv{k} \right)\right|^2
\end{flalign}

\noindent where $\Pi_{j,j'}\left(\bv{k} \right) = \expn{i\bv{k}\left(\bv{s}_j - \bv{s}_{j'}\right)} \sum_{\bar{\bv{R}}} \expn{i\bv{k}\bv{\bar{R}}} \bra{\phi_{-\bar{\bv{R}},j'}} \bv{p} \ket{\phi_{0,j}}$ with summation over nearest hopping neighbor unit cells denoted by $\bv{\bar{R}}$, which is only related to the wavevector, orbital position and the momentum matrix element between two localized atomic orbitals.

The low energy Hamiltonian is written as:

\begin{flalign*}
H(q)=
  \left[
  \begin{array}{ c c c c}
    E  & i2\delta t_1 - qat_1  & 2t_2 & 0 \\
    -i2\delta t_1 - qat_1  & -E  & 0 & 2t_2\\
    2t_2  & 0  & -E & i2\delta t_1 - qat_1\\
    0  & 2t_2  & -i2\delta t_1 - qat_1 & E
  \end{array} 
  \right]
\end{flalign*}

\noindent with respect to the four orbitals shown in Figure~\ref{tb_bands}. Based on this Hamiltonian, the band edge states can be solved as Equations 4 and 5.

When calculating the transition intensity for band edge transitions, the transition intensity can be further simplified as:

\begin{flalign}
 \mathcal{I}(q) &= \left|C_{j'=1,q}^{v \ast} C_{j=0,q}^{c}\Pi_{j=0,j'=1} + C_{j'=2,q}^{v \ast} C_{j=3,q}^{c}\Pi_{j=3,j'=2}\right|^2 \nonumber\\
                     &= \left|\expn{i\theta}+\expn{-i\theta}\right|^2\left|\Pi\left(q\right)\right|^2 \nonumber\\
                     &\equiv \left|W^{v,c}\left(q\right)\right|^2\left|\Pi\left(q\right)\right|^2
\end{flalign}

\noindent In this model, $\Pi_{j=0,j'=1}=\Pi_{j=3,j'=2}$. The transition intensity is only related to the wavefunction coefficient $C$ and the wavefunction phase mismatch between two neighboring chains.

%\bibliography{rappecites}

\end{document}